\documentclass[aps,prc,twocolumn,superscriptaddress]{revtex4-2}
\usepackage{xcolor}
\usepackage{graphicx}
\usepackage{subfig}    
\usepackage{overpic}
\usepackage{caption}
\usepackage{booktabs} 

\begin{document}


\title{Pseudorapidity density distributions of charged
particles and transverse momentum spectra of identified particles in pp collisions in PACIAE 4.0 model}


\author{Zhen Xie}
\affiliation{School of Physics and Information Technology, Shaanxi Normal University, Xi'an 710119, China}


\author{An-Ke Lei}
\affiliation{Key Laboratory of Quark and Lepton Physics (MOE) and Institute of Particle Physics, Central China Normal University, Wuhan 430079, China}

\author{Hua Zheng}
\email[]{zhengh@snnu.edu.cn}

\author{Wenchao Zhang}
\affiliation{School of Physics and Information Technology, Shaanxi Normal University, Xi'an 710119, China}

\author{Dai-Mei Zhou}
\email[]{zhoudm@mail.ccnu.edu.cn}
\affiliation{Key Laboratory of Quark and Lepton Physics (MOE) and Institute of Particle Physics, Central China Normal University, Wuhan 430079, China}

\author{Zhi-Lei She}
\affiliation{School of Mathematical and Physical Sciences, Wuhan Textile University, Wuhan 430200, China}

\author{Yu-Liang Yan}
\affiliation{China Institute of Atomic Energy, P. O. Box 275 (10), Beijing 102413, China}

\author{Ben-Hao Sa}
\email[]{sabhliuym35@qq.com}
\affiliation{Key Laboratory of Quark and Lepton Physics (MOE) and Institute of Particle Physics, Central China Normal University, Wuhan 430079, China}
\affiliation{China Institute of Atomic Energy, P. O. Box 275 (10), Beijing 102413, China}


\date{\today}

\begin{abstract}
The pseudorapidity density distributions of charged particles and the transverse momentum spectra of identified particles in proton-proton (pp) collisions at the center-of-mass energies ranging from $\sqrt{s}=200$ GeV to 13 TeV have been systematically studied using the newly released parton and cascade model PACIAE 4.0 based on PYTHIA 8.3. The available experimental data are well reproduced across all analyzed aspects. This theoretical method can be easily extended to 
anywhere the experimental data for pp collisions are currently unavailable. Furthermore, since pp collisions serve as the baseline for heavy-ion collisions, our 
results can provide a valuable resource for both experimentalists and 
theorists. 
\end{abstract}


\maketitle

\section{INTRODUCTION}{\label{intro}}
The main aim of the high energy proton-proton (pp) and heavy-ion collision experiments is to explore the properties of a novel form of matter termed as quark-gluon plasma (QGP) \cite{STAR:2005gfr,PHOBOS:2004zne,Tan:2024lrp,Chen:2024aom,qgp3,qgp4,Zhao:2020wcd,Zhao:2021vmu}. Here the basic observables are the particle yields, collective flows, pseudorapidity density distributions of charged (identified) particles, and the particle transverse momentum spectra etc.. These observables,  depending on the collision system, the collision centrality, and the collision energy, have been extensively 
investigated both experimentally and theoretically~\cite{PHOBOS:2010eyu,STAR:2017sal,Toia:2011nzq,Deppman:2019yno,Shi:2024pyz,Tao:2023kcu,Tao:2022tcw,Zhu:2022dlc,Zhu:2022bpe,Tao:2020uzw,Zhu:2021fbs,Zheng:2015mhz,Wang:2022det,Zhu:2018nev,Gao:2017yas,Zheng:2015gaa,Zheng:2015tua,Wong:2015mba,Rath:2019cpe,Xu:2017akx,Yang:2022fcj}. They serve 
as the probes for understanding the physics processes during the collisions, including particle production mechanisms. 

Meanwhile, the pseudorapidity density distributions of charged particles ($\frac{dN_{ch}}{d\eta}$) and the transverse momentum spectra of identified particles in pp collisions are important benchmark tools to constrain the event generators (models) among PYTHIA~\cite{Sjostrand:2006za,Bierlich:2022pfr}, 
HERWIG \cite{Corcella:2000bw}, HIJING \cite{Wang:1991hta}, 
AMPT \cite{Lin:2004en}, SMASH \cite{PhysRevC.94.054905}, PACIAE \cite{Lei:2023srp,Lei:2024kam}, etc., and to provide insights into various physics searches in the collisions. Their measurements are essential in characterizing the particle production mechanisms and the formation of QGP.

The international collaborations have measured a substantial amount of
experimental data of $\frac{dN}{d\eta}$ and particle transverse momentum
spectra~\cite{PHOBOS:2010eyu,STAR:2006xud,STAR:2006nmo,ALICE:2013jfw,ALICE:2015olq,ALICE:2019hno,ALICE:2014juv,ALICE:2015ial,ALICE:2015qqj,ALICE:2010cin,ALICE:2011gmo,ALICE:2022kol} 
ranging from small to large collision systems at various collision centralities and energies. The pseudorapidity density distributions of charged particles and the transverse momentum spectra of charged (identified) particles are generally limited to 
the midrapidity region, due to the advantageous kinematic 
conditions in the midrapidity and
the experimental constraints encountered in the forward and backward rapidity regions.  
This motivates us to search for particle production in full pseudorapidity ranges theorectically. 

\begin{figure*}
\raggedright
\includegraphics[width=1.0\textwidth]{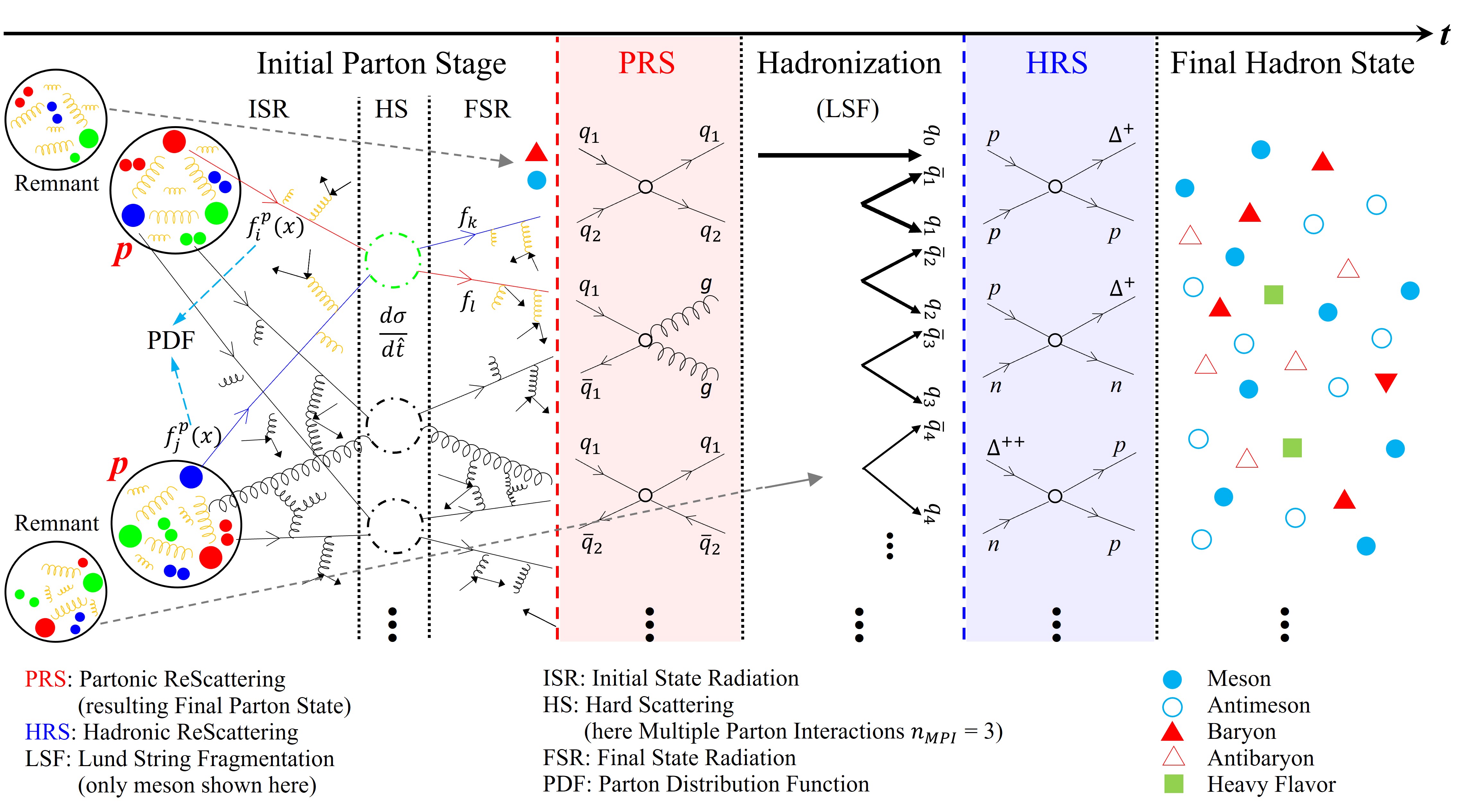}
\caption{The physics processes develope continuously in a high energy pp collision in PACIAE 4.0.}
\label{fig0}
\end{figure*}

\begin{figure}
\raggedright
\includegraphics[width=0.48\textwidth]{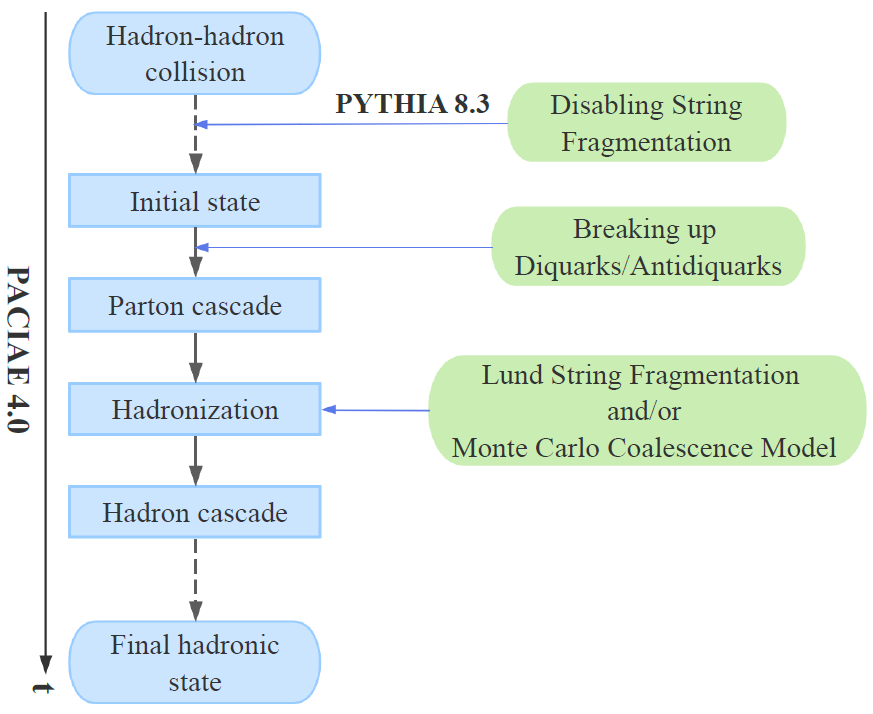}
\caption{ The program flow of a pp collision in PACIAE 4.0.}
\label{fig1}
\end{figure}

In this paper, we perform detailed simulations for the pp collisions with 
the newly released parton and hadron cascade model PACIAE 4.0 \cite{Lei:2024kam}. The PACIAE 
is a phenomenological model Monte Carlo event generator developed to describe 
the high-energy lepton-lepton, lepton-hadron, lepton-nucleus, hadron-hadron, 
hadron-nucleus, and nucleus-nucleus collisions based on the PYTHIA model. It 
can switch flexibly between PYTHIA 6 \cite{Sjostrand:2006za} and PYTHIA 8 \cite{Bierlich:2022pfr}. The main difference between 
PACIAE and PYTHIA for pp collisions is introducing the parton rescattering and the hadron 
rescattering before and after hadronization, respectively~\cite{Lei:2023srp}. Our study aims 
to systematically simulate the pseudorapidity density distributions of charged particles and the transverse momentum spectra of identified particles in pp collisions at center-of-mass energies ranging from 
$\sqrt{s}= 200$ GeV to 13 TeV and to compare them with the corresponding 
experimental data. The properly fitted parameters and switches provide a 
standard tune for both experimentalists and 
theorists.

The paper is organized as follows. In Sec. \ref{mod}, we briefly introduce 
PACIAE 4.0 model. Section \ref{rad} shows the simulated results of the pseudorapidity density distributions of charged particles, the transverse momentum spectra of identified particles and the particle ratios of $K/\pi$ and $p/\pi$ in pp collisions at various collision energies and the comparison with experimental data. A brief conclusion 
is presented in Sec. \ref{con}.
 
\section{The parton and hadron cascade model PACIAE 4.0}\label{mod}

The PACIAE model is a phenomenological model Monte Carlo event generator. It is capable of the simulation for the high-energy elementary particle collisions and 
nucleus-nucleus collisions. The latest release, PACIAE 4.0, is based on PYTHIA 8.3 instead of PYTHIA 6.4 originally. A brief introduction to PACIAE 4.0 has been recently published in Ref.~\cite{Lei:2024kam}, where the model structure, program flow, and all the new features compared to earlier versions based on PYTHIA 6.4 are detailed. Here we only emphasize the physics aspects relevant to our anaysis as follows.

For the pp collisions, the PACIAE 4.0 simulation comprises four stages: the initial partonic state stage, the partonic rescattering (PRS) stage, the hadronization stage, and the hadronic rescattering (HRS) stage. The initial state is prepared by running PYTHIA 8.3 with temporary disabling the Lund string fragmentation and the postsetting of breaking up diquarks/anitdiquarks randomly. By default, PYTHIA does not provide the four-coordinate information of partons, i.e., the four-coordinate of each parton is $(0,0,0,0)$. In PACIAE, we randomly distribute the initial partons on a spherical surface with unit-radius (which is comparable to the radius of a proton) with origin of pp collision vertex. The formation times of the initial partons are assumed to be equal to zero, as they are expected to be small. This initial parton state then undergoes $2\rightarrow 2$ partonic rescattering using the lowest order Perturbative Quantum Chromodynamics (pQCD) massive cross section. The string configurations keep no change after elastic partonic scatterings, assumed in this work, while the positions and momenta of the inner partons are modified by the scatterings. The resulted final partonic state is then hadronized by the Lund string fragmentation mechanism in this work. We then distribute the hadrons from Lund string fragmentation randomly on a spherical surface with radius 1.16 fm (approximately corresponding to the typical radius of hadrons) centered on a randomly selected string endpoint from the corresponding string. Finally, the resulted hadrons undergo $2\rightarrow 2$ hadronic rescattering with empirical cross sections, providing the final hadronic state. These four stages are displayed vividly in the Fig.~\ref{fig0}. The Fig.~\ref{fig1} is just the corresponding 
block diagram for a pp collision in PACIAE 4.0.

There are two pivotal parameters $a$ and $b$ relevant to the Lund string fragmentation probability \cite{Andersson:1983ia,Sjostrand:2006za}:
\begin{equation}
    f(z) \propto (1/z) \cdot (1-z)^a \cdot \exp(-b m_T^2 / z),
\end{equation}
which is associated with particle production in the simulation. In the above equation, $z$ is the energy fraction taken away by a hadron fragmented from a high energy parton, $m_T=\sqrt{m^2+p_T^2}$ is the transverse mass of the hadron. In PACIAE 4.0, $a=0.68 $ is the default value which is adopted and fixed in this work. However the parameter $b$ is left to fit the pseudorapidity density distributions of charged particles and the transverse momentum spectra of identified particles simultaneously in pp collisions at a given collision energy, which serves as our parameter adjustment strategy. We aim to provide a reference parameter set for the users and keep it as simple as possible.
\begin{figure}
\raggedright
\includegraphics[width=0.48\textwidth]{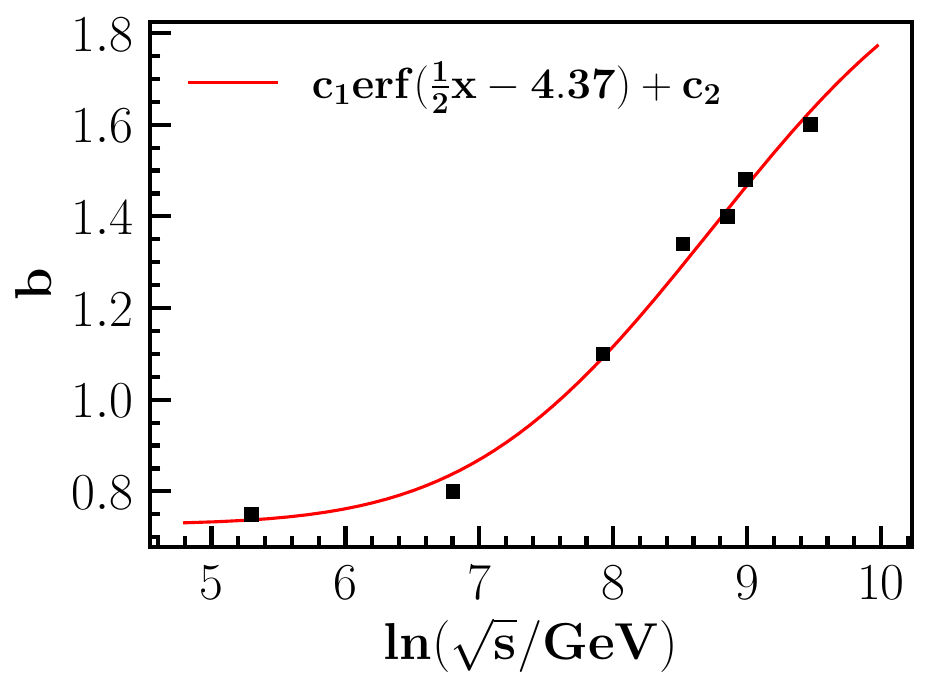}
\caption{The parameter $b$ versus the pp collision energy. The line is the fit with the function in the legend where $x=\ln(\sqrt{s}/{\rm GeV})$. }
\label{fig6}
\end{figure}

\begin{table}[ht]
	\centering
	\caption{Values of parameter $b$ for various collision energies in pp collisions for PACIAE 4.0 model simulations when $a=0.68$ and $K=0.8$ are fixed.}
	\renewcommand{\arraystretch}{1.5} \tabcolsep 20pt%
	\begin{tabular*}{0.35\textwidth}{cc}
    \toprule
    \hline
    $\sqrt{s}$  & $b$ (adj1(7))  \\\hline
        200  GeV & 0.75 \\
        0.9  TeV & 0.80\\
	  2.76 TeV & 1.1\\
        5.02 TeV &1.34 \\
         7   TeV & 1.4\\
         8   TeV & 1.48\\
        13   TeV & 1.6\\
   \hline
   \bottomrule
	\end{tabular*}\label{tab1}
	\end{table}
    
\begin{figure}
\raggedright
\includegraphics[width=0.48\textwidth]{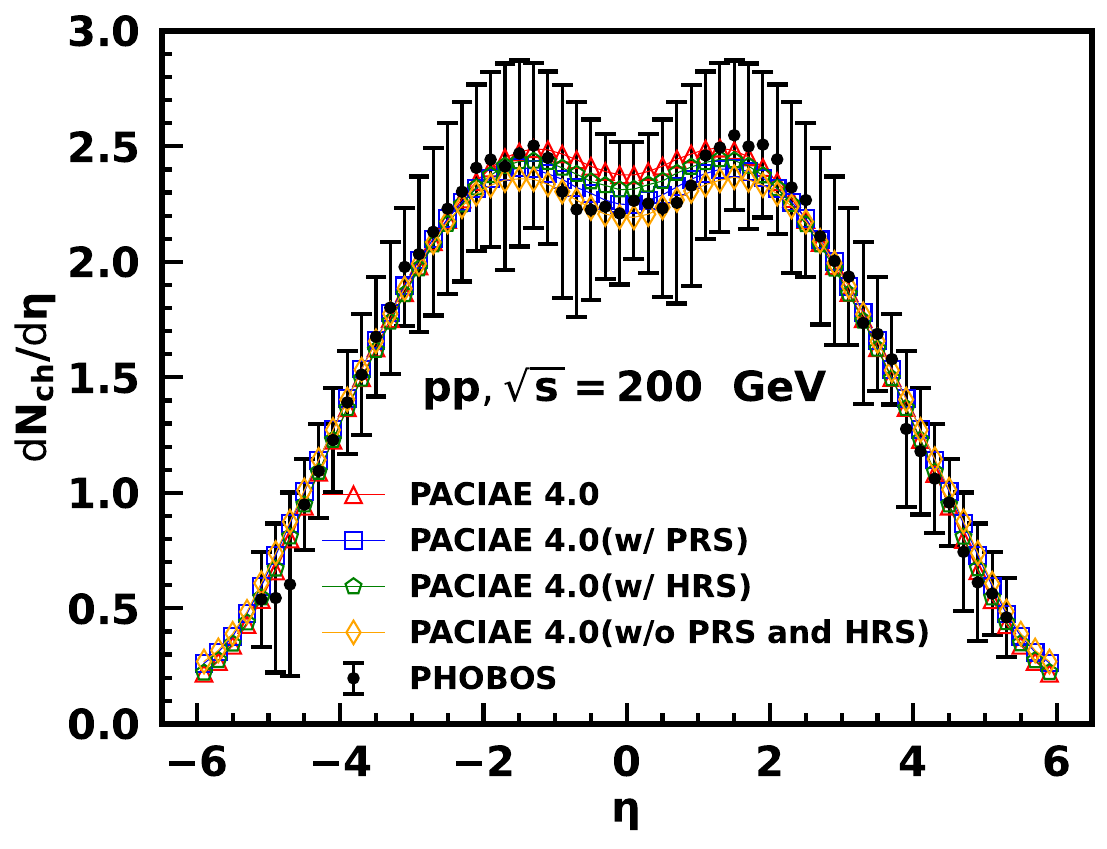}%
\caption{ The pseudorapidity density distribution of charged particles produced in pp collisions at $\sqrt{s}= 200$ GeV from PACIAE 4.0 simulations compared with experimental data from PHOBOS \cite{PHOBOS:2010eyu}. }
\label{fig2}
\end{figure}

\begin{figure}[!ht]
   \centering
   \includegraphics[width=0.45\textwidth]{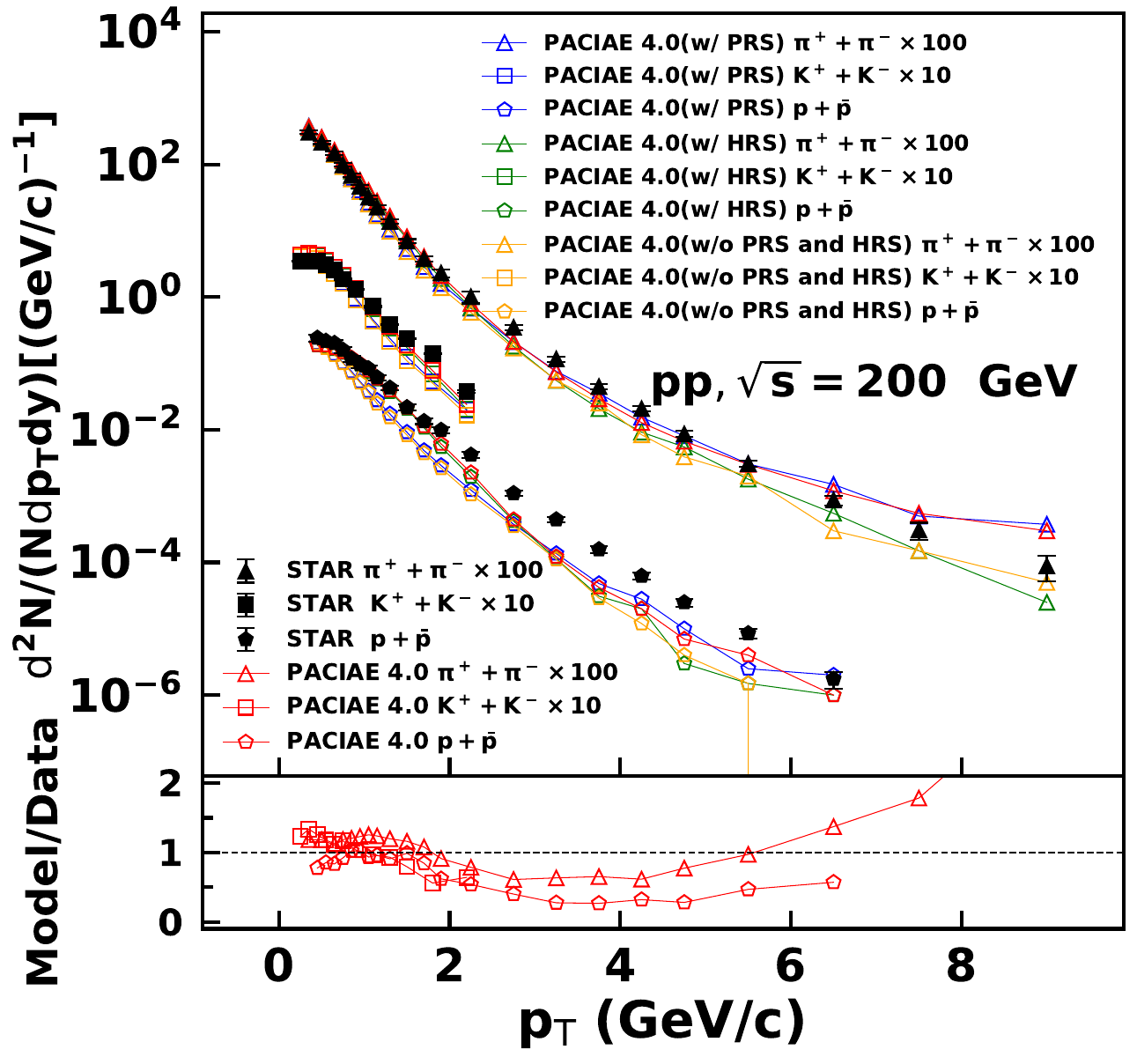}%
   \caption{The transverse momentum spectra for $\pi^++\pi^-$, $K^++K^-$, and $p+\bar{p}$  at midrapidity 
 in pp collisions at $\sqrt{s}=$200 GeV from PACIAE simulations compared with STAR data \cite{STAR:2006xud,STAR:2006nmo}.}
  \label{fig3}
\end{figure}

\begin{figure*}[!ht]
   \centering
    \subfloat{
	\begin{overpic}[width=0.32\linewidth]{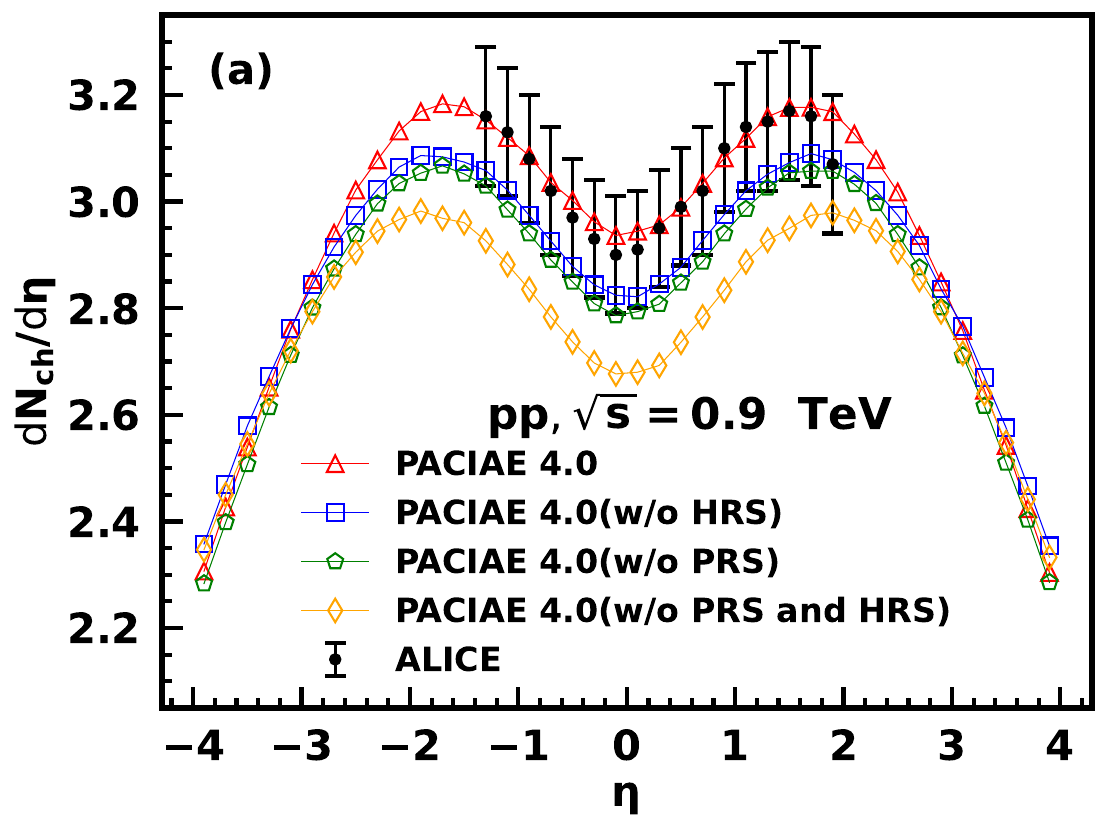}
    \end{overpic}
	}
    \hfill
     \subfloat{
	\begin{overpic}[width=0.32\linewidth]{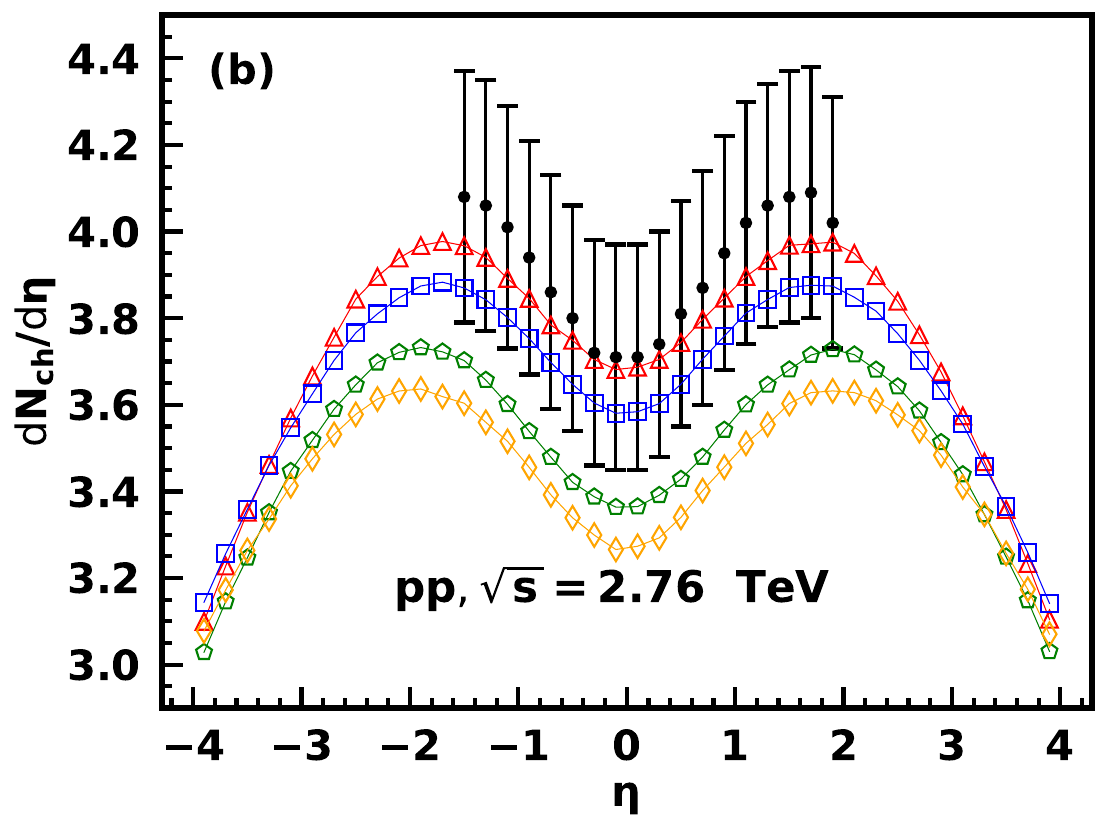}
    \end{overpic}
	}
    \hfill
    \subfloat{
	\begin{overpic}[width=0.32\linewidth]{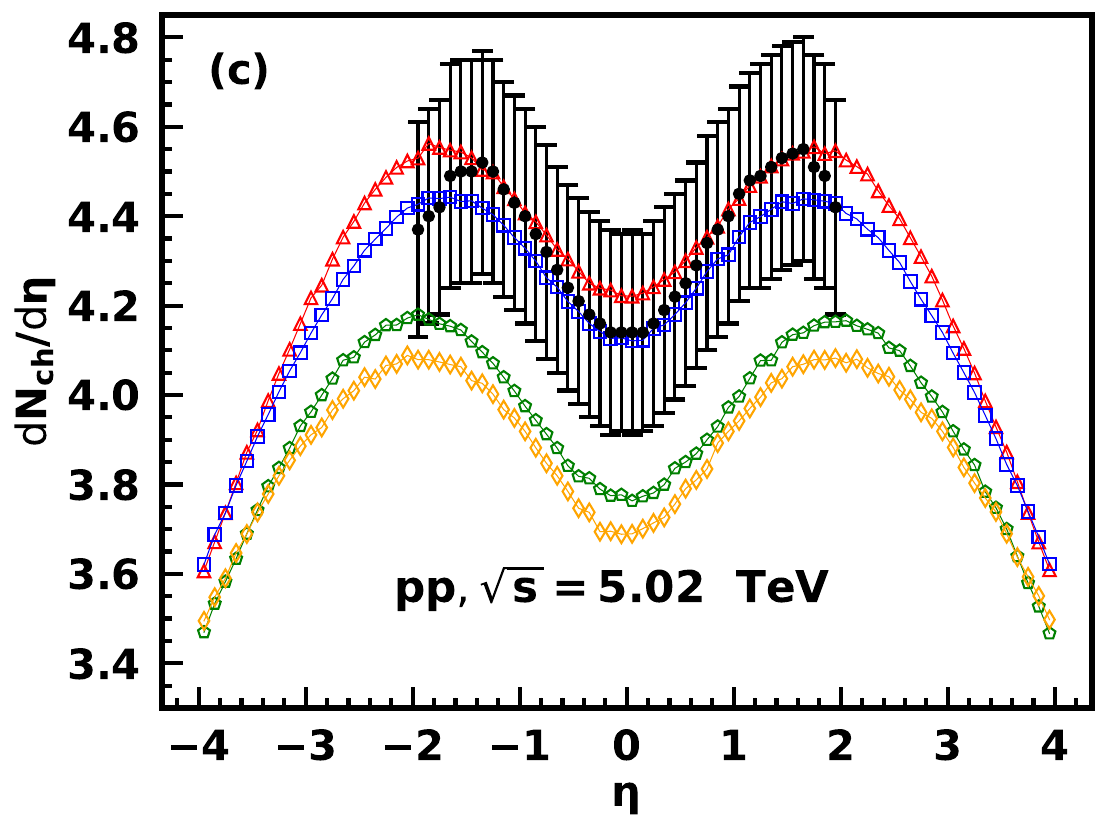}
    \end{overpic}
	}
    \hfill
     \subfloat{
	\begin{overpic}[width=0.32\linewidth]{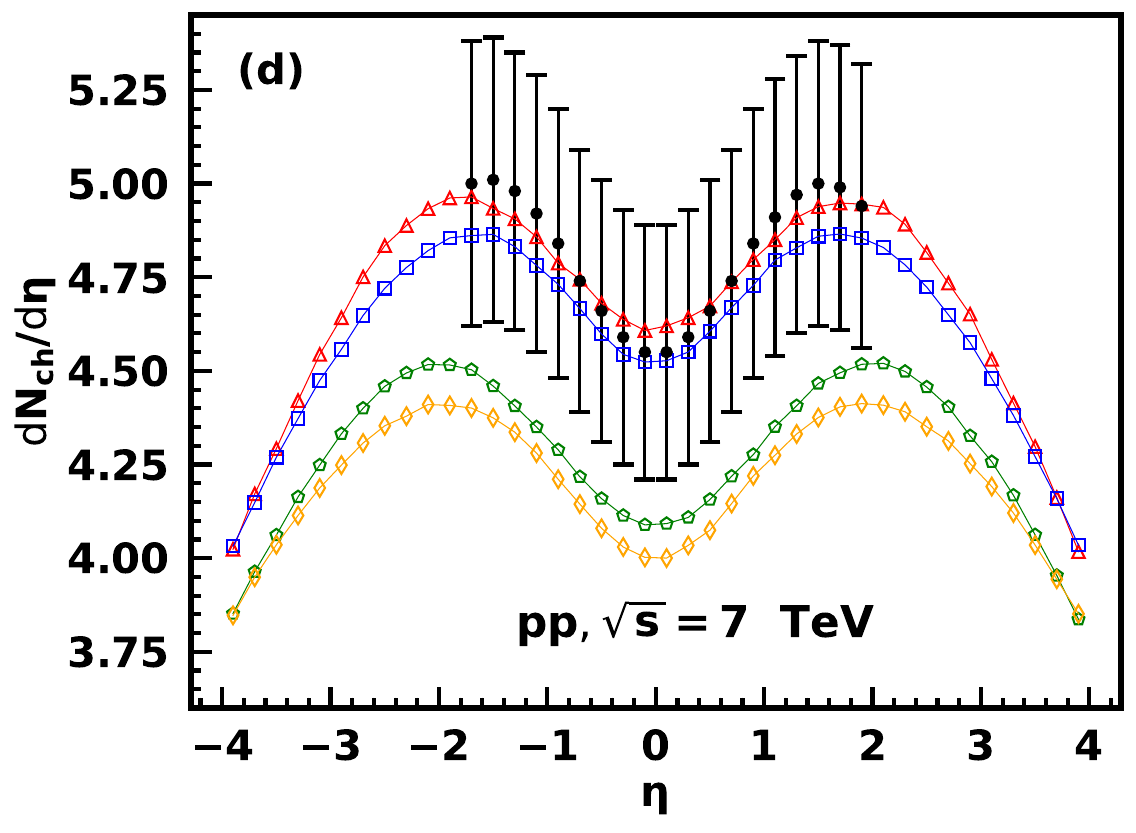}
    \end{overpic}
	}
    \hfill
   \subfloat{
	\begin{overpic}[width=0.32\linewidth]{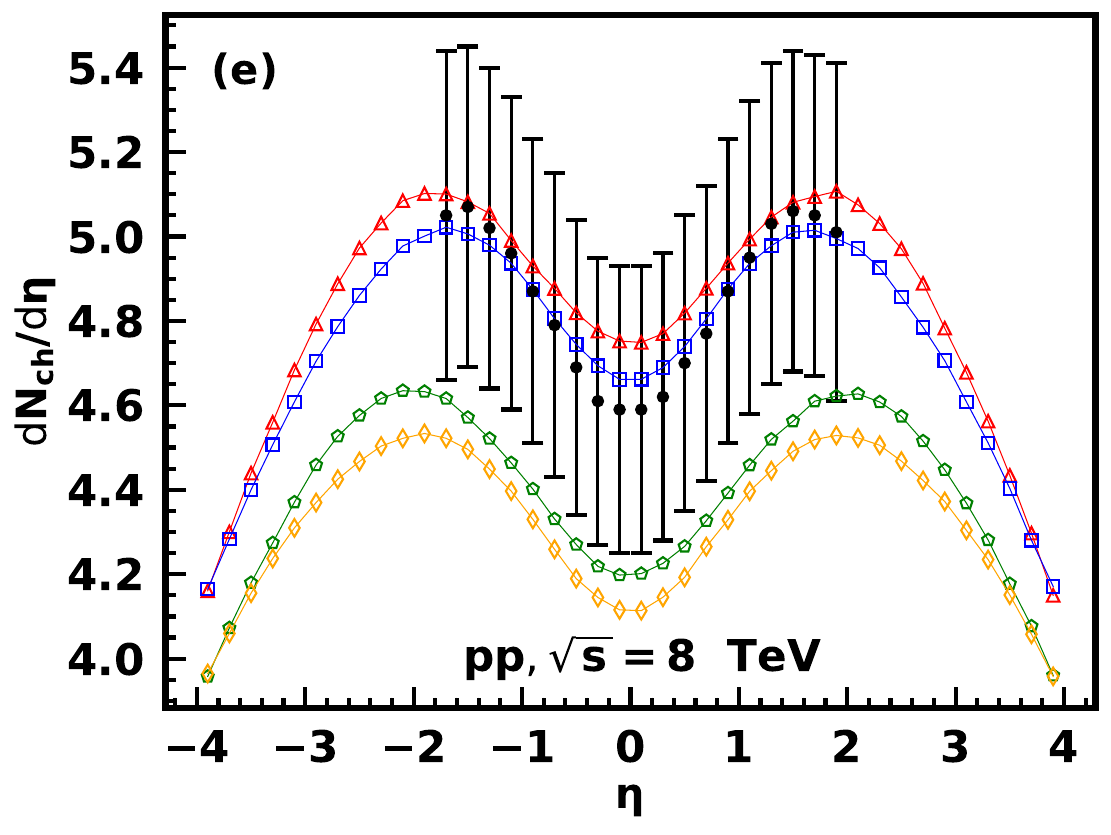}
    \end{overpic}
	}
    \hfill
    \subfloat{
	\begin{overpic}[width=0.32\linewidth]{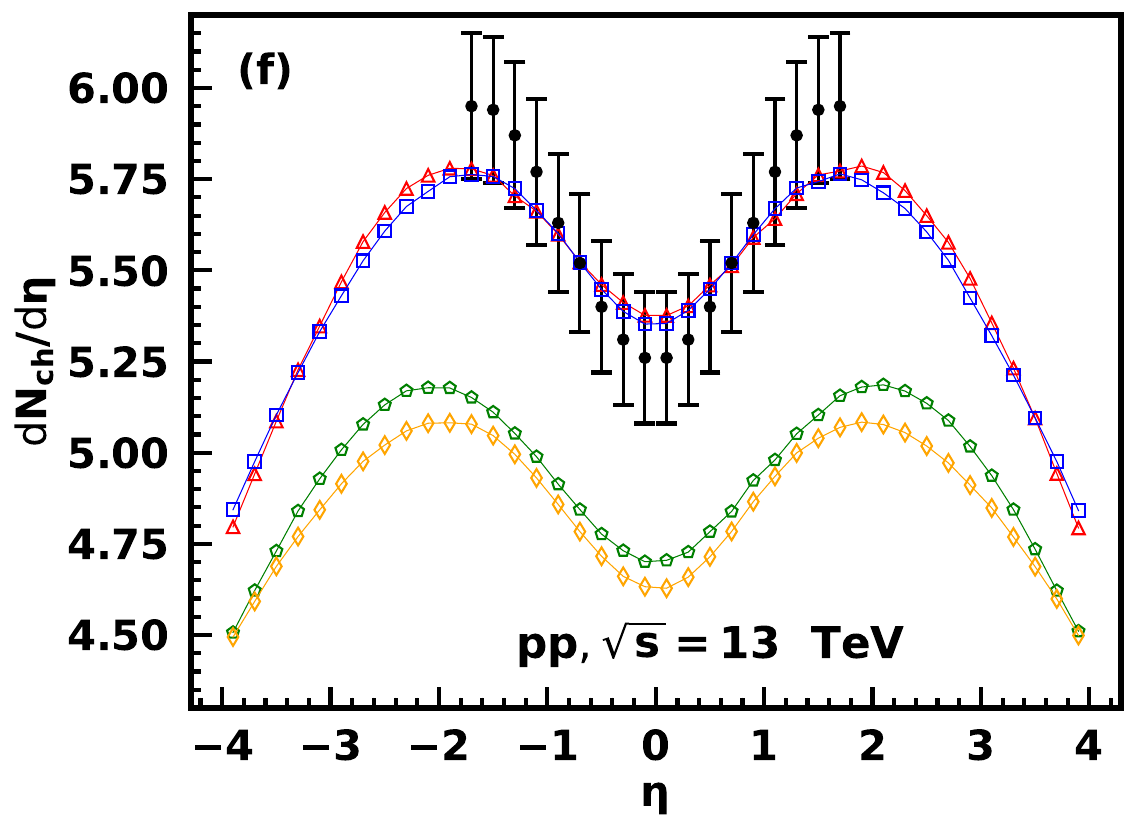}
    \end{overpic}
	}
   \caption{The pseudorapidity density distributions of charged particles produced in pp collisions at $\sqrt{s}= 0.9-13$ TeV from PACIAE 4.0 model simulations compared with ALICE data \cite{ALICE:2015olq,ALICE:2015qqj,ALICE:2022kol}.}
  \label{fig4}
\end{figure*}

Another important parameter is the $K$ factor (adj1(10) in the program), which is a multiplicative factor of the hard scattering cross sections \cite{Combridge:1977dm,Field:1989uq} 
\begin{equation}
    \frac{d\sigma}{dt}(ab\rightarrow cd;s,t)= K\frac{\pi\alpha_s^2}{s^2}|\bar{M}(ab\rightarrow cd)|^2,
\end{equation}
where $\alpha_s$ refers to strong coupling constant. The variables $s, t$ are 
the Mandelstam invariants in the kinematics of the $ab\rightarrow cd$ quark 
process. In our work, $K=0.8$ is fixed  by fitting the experimental data for all collision energies. Generally speaking, larger
$K$ factor, i.e. lager cross sections, leads to larger number of hard scattering processes and
increases the total multiplicity of charged particles. The pseudorapidity density distribution of charged particle is more sensitive to $K$ factor than the transverse momentum spectra of identified particles, because the latter is only collected from the midrapidity region. Here one has to note that the increasing of $a$ parameter and/or dereasing $b$ parameter are also possible to increase total multiplicity of charged particles.

\section{RESULTS AND DISCUSSION }\label{rad}

In a pp collision PACIAE 4.0 simulation, we are able to follow the trajectory and the physics processes for each particle (parton or hadron) and that is just the advantage of the Monte Carlo method. Therefore, all the information of particles are recorded and can be extracted. These information enable us to learn the physics that occured in the evolution processes of the collision system. 

As a high energy event generator, it should reproduce as many experimental observables as possible to validate the capability of the model, which is called calibration in the experimental terminology. The experimental data of the pseudorapidity density distributions of charged particles and the transverse momentum spectra of the majority identified charged particles of $\pi^++\pi^-$, $K^++K^-$, and $p+\bar{p}$ are fitted by the simulated results, respectively, in the pp collisions at various collision energies. The former determines the total multiplicity of charged particles produced and the latter shows the capability of the model to reproduce the detailed distributions of the final particles in the pp collisions at a given collision energy. The fitted values of parameter $b$ are listed in Table~\ref{tab1} which are provided as references for the users. Figure \ref{fig6} illustrates  the variation of the parameter $b$ in PACIAE 4.0 model across different collision energies while parameters $a$ and $K$ factor are fixed. The red line is the fit with the function displayed in the legend which can be used to estimate the value of $b$ for other collision energies. The functional form is purely phenomenological with the principle that we use as few free parameters as possible. It should also be able to describe the parameter trend across the energy range.

\begin{figure*}[!ht]
  \raggedright
    \subfloat{
	\begin{overpic}[width=0.42\linewidth]{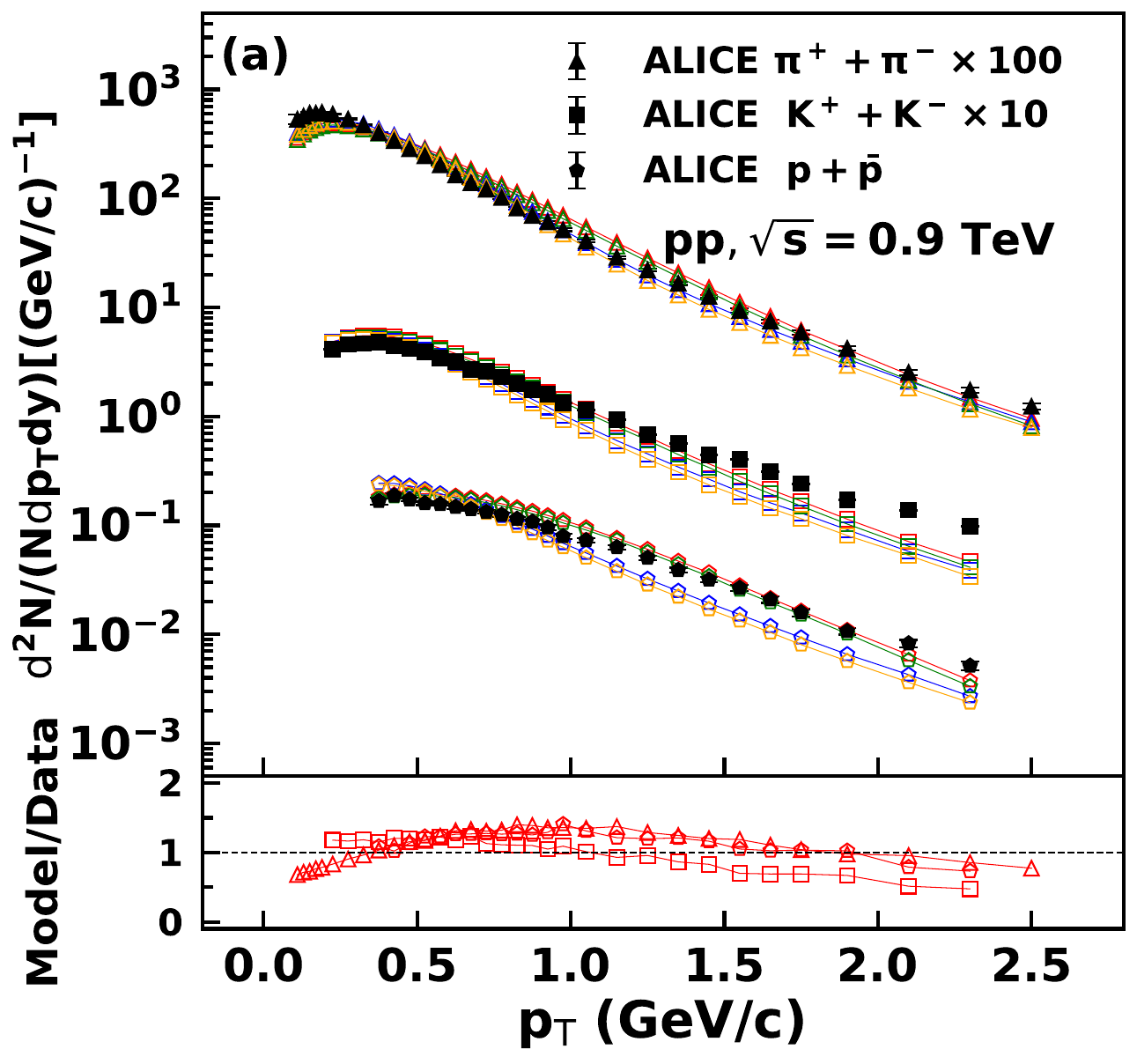}
    \end{overpic}
	}
    \hfill
     \subfloat{
	\begin{overpic}[width=0.42\linewidth]{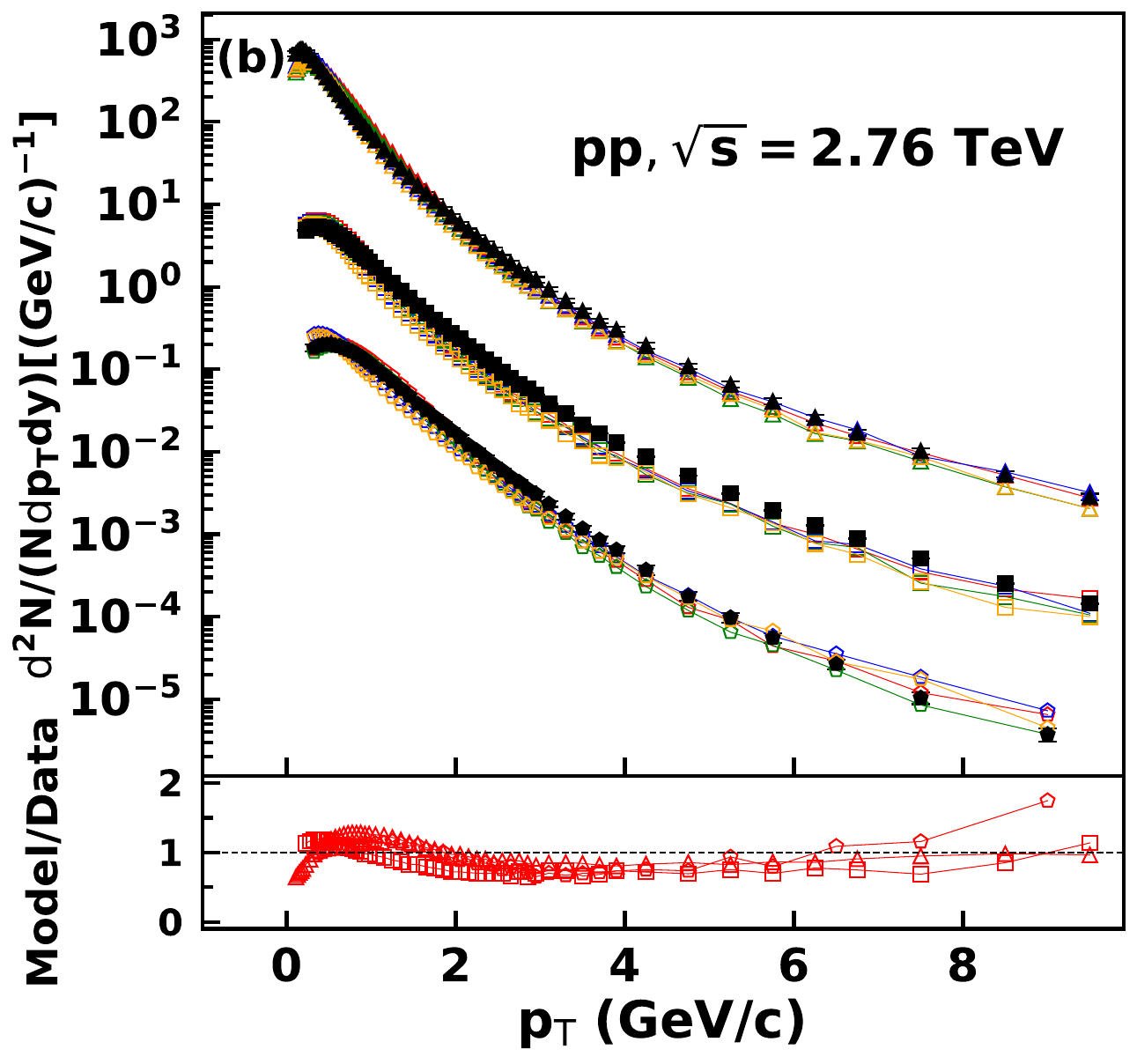}
    \end{overpic}
	}
    \hfill
    \subfloat{
	\begin{overpic}[width=0.42\linewidth]{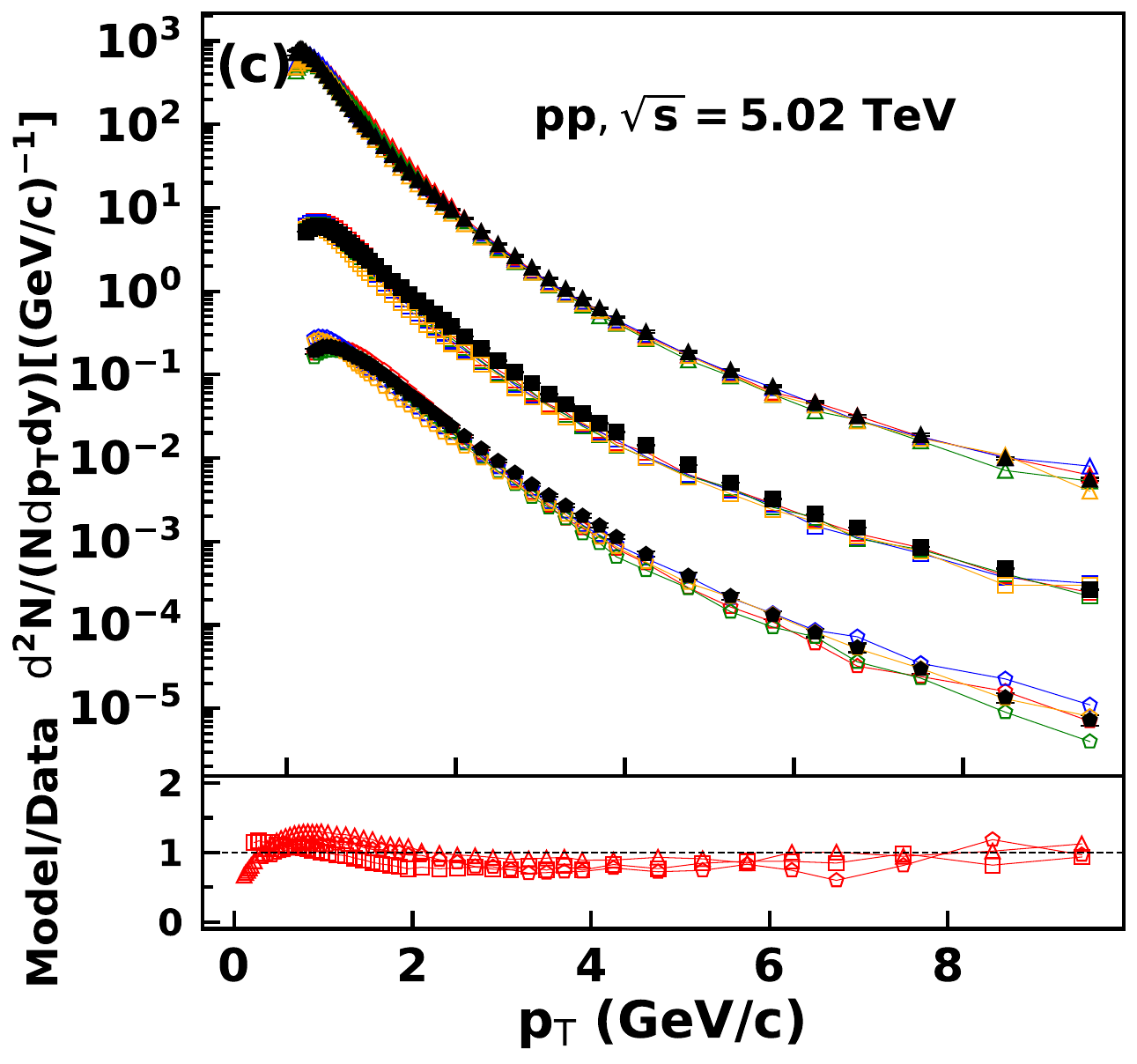}
    \end{overpic}
	}
    \hfill
    \subfloat{
	\begin{overpic}[width=0.42\linewidth]{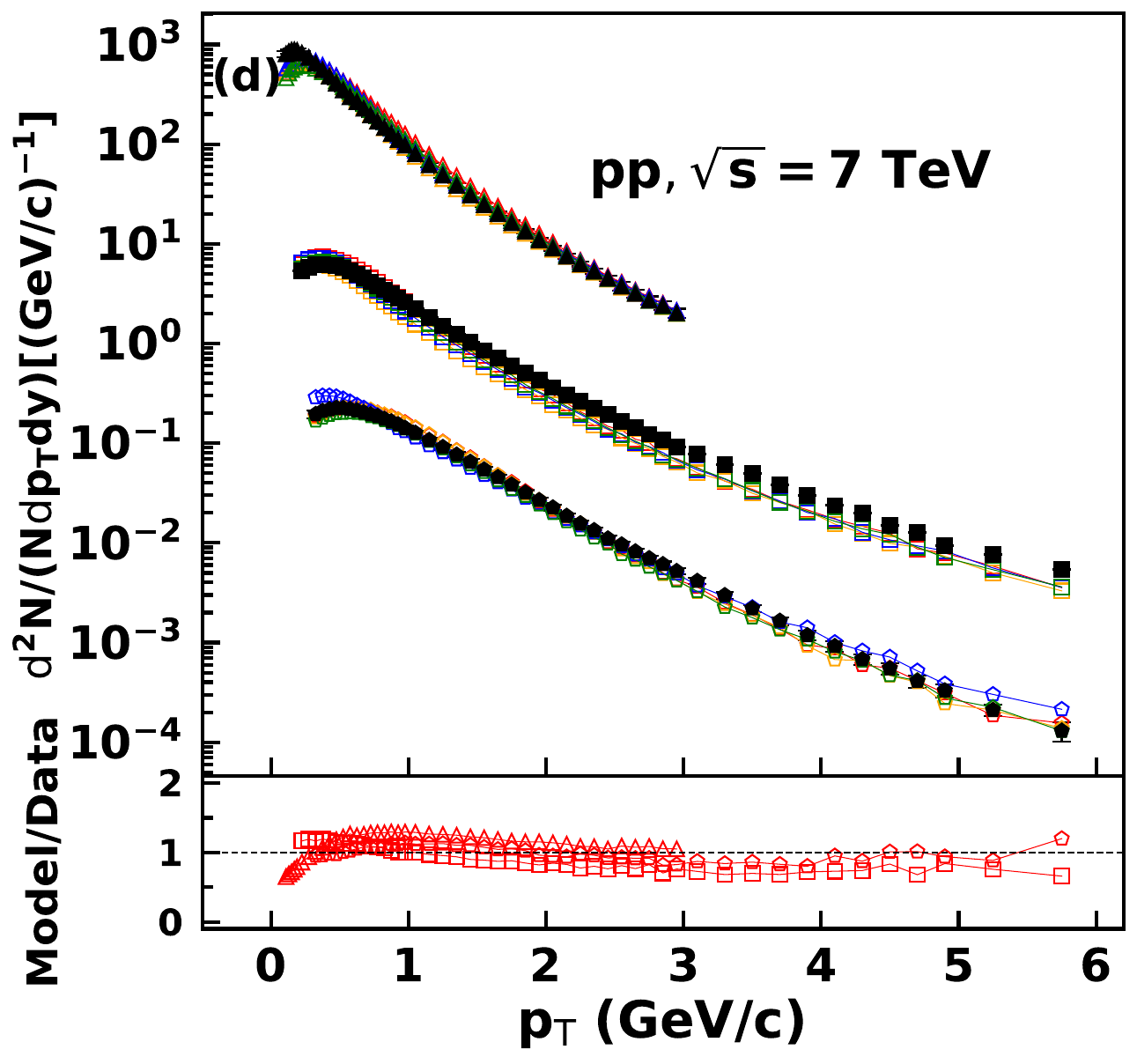}
    \end{overpic}
	}
    \hfill
   \subfloat{
	\begin{overpic}[width=0.98\linewidth]{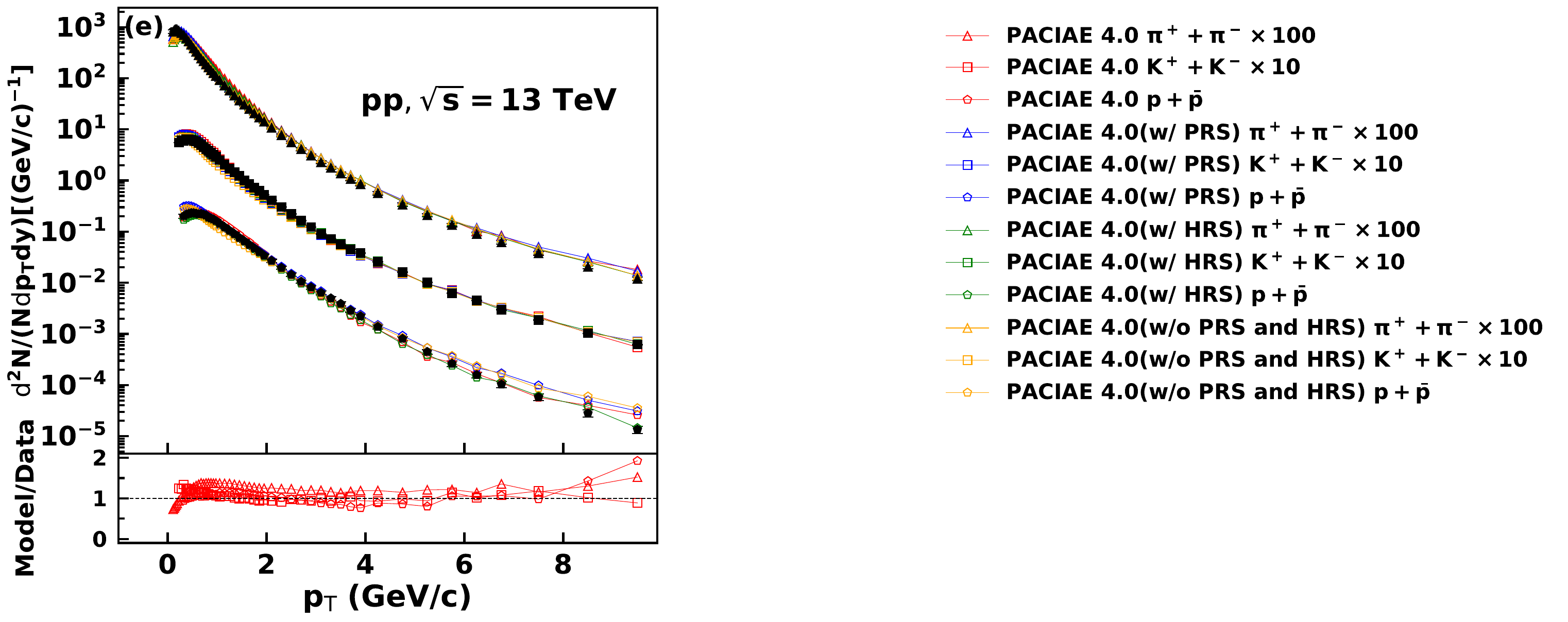}
    \end{overpic}
	}

   \caption{The transverse momentum spectra for $\pi^-+\pi^+$, $K^++K^-$, and $p+\bar{p}$ at midrapidity
 in pp collisions at $\sqrt{s}= 0.9-13$ TeV from PACIAE 4.0 model simulations compared with ALICE data \cite{ALICE:2019hno,ALICE:2014juv,ALICE:2015ial,ALICE:2011gmo}.}
  \label{fig5}
\end{figure*}

Figure \ref{fig2} shows the pseudorapidity density distribution of charged particles in pp collisions at $\sqrt{s}=200$ GeV. Notice that `w/' and `w/o' are abbreviations for `with' and `without' in the legend, respectively (hereinafter similarly). The black dots with errors represent the experimental data from PHOBOS collaboration at RHIC \cite{PHOBOS:2010eyu}. The red open triangles are the simulation results with both partonic rescattering and hadronic rescattering. The blue open squares are for the case with only partonic rescattering. The green open pentagons are for the case with only hadronic rescattering. The yellow open diamonds correspond to case without partonic rescattering and hadronic rescattering, which are the results from PYTHIA 8.3. It is observed that the simulated results for the four cases reproduce the experimental pseudorapidity density distribution of charged particles well in the whole rapidity region, indicating that the rescattering effects are weak for the total charged particles yield in this relative low collision energy referring to the collision energies at LHC. It can be understood because there are not so many partons and hadrons, resulting in not so many rescatterings occurring. It will be shown that both partonic rescattering and hadronic rescattering can increase the total multiplicity of charged particles at higher collision energies later. This is due to the momentum exchange that occurs between the re-scattered particles, which leads to a redistribution of the energy and momentum, increasing the multiplicity of particles. The distribution is symmetric referring to $\eta =0$ and decreases at the forward/backward rapidity region as expected. 

Figure \ref{fig3} presents the transverse momentum spectra of identified particles ($\pi^++\pi^-$, $K^++K^-$, and $p+\bar{p}$) at midrapidity in pp collisions at $\sqrt{s}=200$ GeV. The simulated results (red open symbols) are compared with the corresponding experimental data (black solid symbols) from the STAR collaboration at RHIC. To better visualize the difference between the simulated results and the experimental data, we also provide the ratios of Model/Data for identified particles. The results show excellent agreement for pions and kaons. There are some derivations for protons at $p_{\rm T}>2$ GeV/c. It is observed that there is interplay between the partonic rescattering effect and the hadronic rescattering effect, which modifies the detailed shapes of the particle spectra to some extent, bringing them closer to experimental data. It demonstrates the importance of the rescattering effects even though their effects to pseudorapidity distribution of charged particles are negligible at this collision energy. The results validate the model's ability to describe the detailed dynamics of particle production and the underlying physics of hadronization. We emphasize that the pseudorapidity distribution of charged particles in Fig. \ref{fig2} and the transverse momentum spectra of identified particles in Fig. \ref{fig3} are reproduced simultaneously in PACIAE 4.0.

Similarly to Fig. \ref{fig2}, Fig. \ref{fig4} shows the pseudorapidity density distributions of charged particles in pp collisions at various collision energies, ranging from $\sqrt{s}=$ 0.9 TeV to 13 TeV at LHC. For all collision energies, the agreement between the model results (red open triangles) and the experimental data is excellent. One can see that both partonic rescattering and hadronic rescattering can increase the multiplicity of charged particles at higher collision energies because the rescattering effects play an important role, as explained before. As the collision energy increases, the partonic rescattering effect becomes dominant because of the large number of partons created. One also can observe that the pseudorapidity density of charged particles increases at midrapidity, reflecting the higher multiplicity of charged particles produced at higher collision energies. The model also accurately captures the broadening of the pseudorapidity distribution at each collision energy.

Similarly to Fig. \ref{fig3}, Fig. \ref{fig5} shows the transverse momentum spectra of identified particles ($\pi^++\pi^-$, $K^++K^-$, and $p+\bar{p}$) produced at midrapidity as well as the ratios of Model/Data in pp collisions at various collision energies, ranging from $\sqrt{s}=$ 0.9 TeV to 13 TeV. The simulated results agree well with the experimental data in the entire $p_{\rm T}$ range for all collision energies. The maximum $p_{\rm T}$ is cut at 10 GeV/c. The interplay between the partonic rescattering effect and hadronic rescattering effect reshapes the details of the particle spectra to some extent, bringing them closer to experimental data across all collision energies. Again, the results in Fig. \ref{fig4} and Fig. \ref{fig5} are reproduced simultaneously with the same set of model parameters at a given collision energy. These results demonstrate its ability to describe the particle production in pp collisions over a wide range of collision energies.

\begin{figure*}[!ht]
  \raggedright
    \subfloat{
	\begin{overpic}[width=0.32\linewidth]{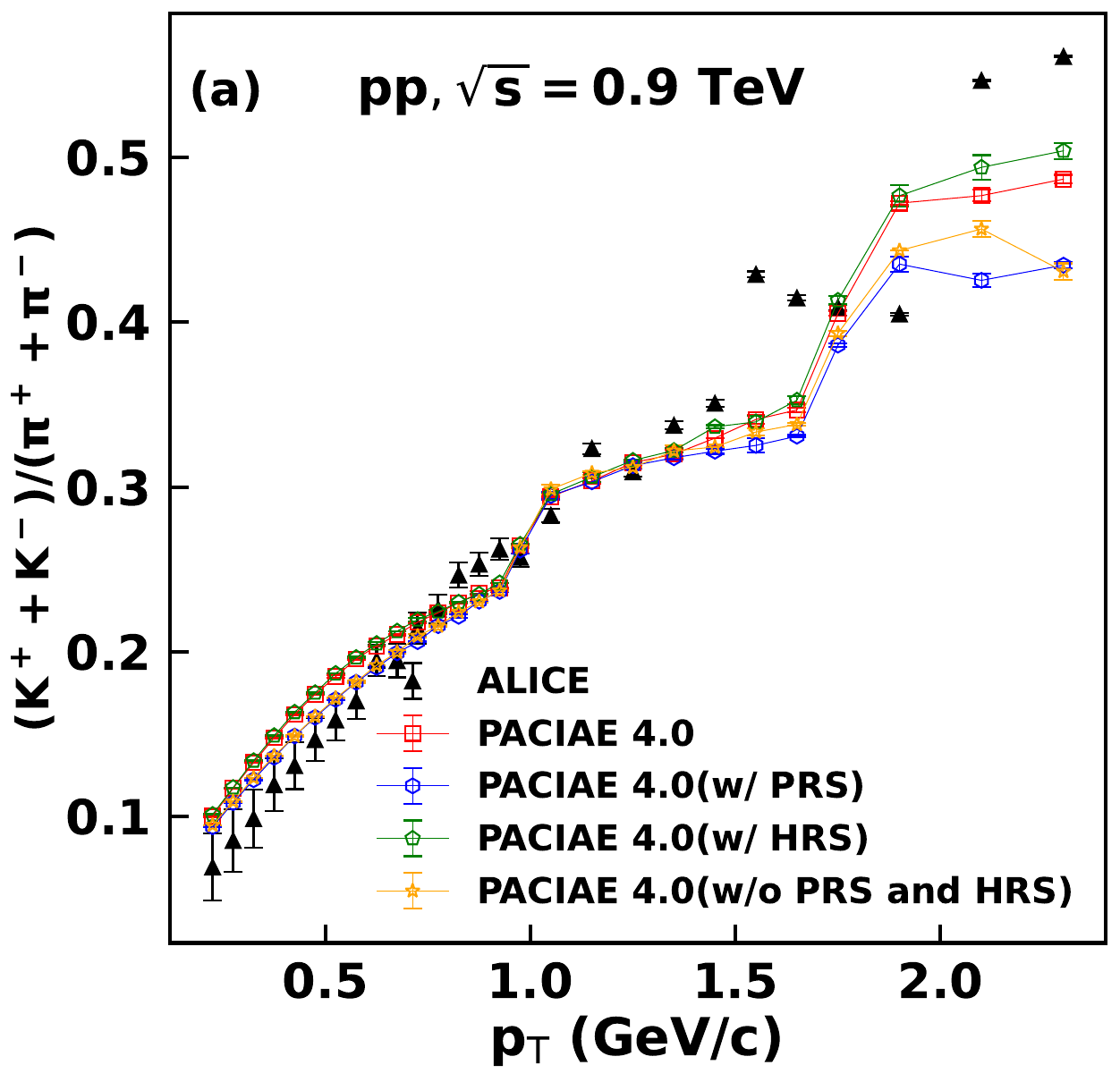}
    \end{overpic}
	}
    \hfill
     \subfloat{
	\begin{overpic}[width=0.32\linewidth]{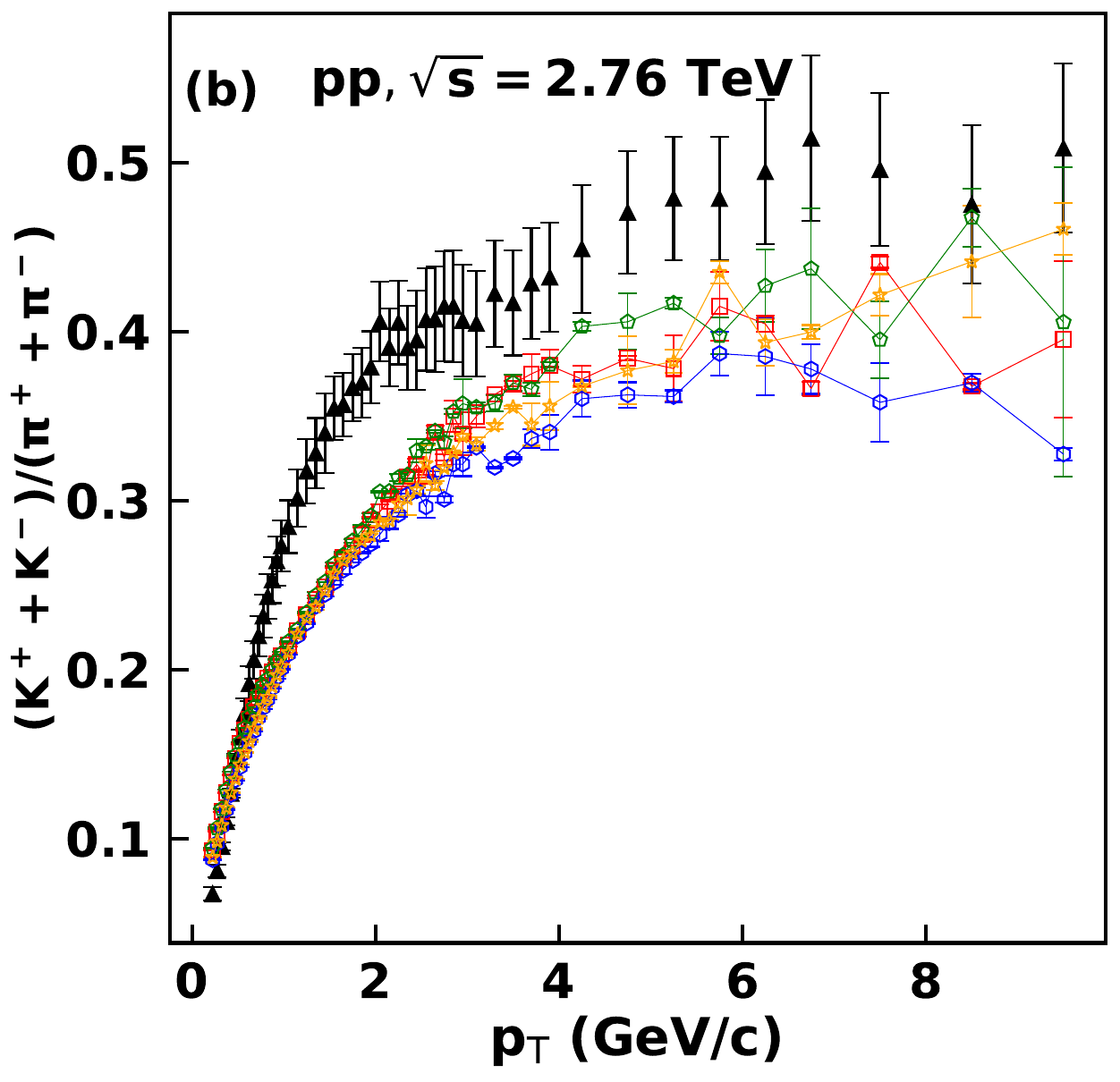}
    \end{overpic}
	}
    \hfill
    \subfloat{
	\begin{overpic}[width=0.32\linewidth]{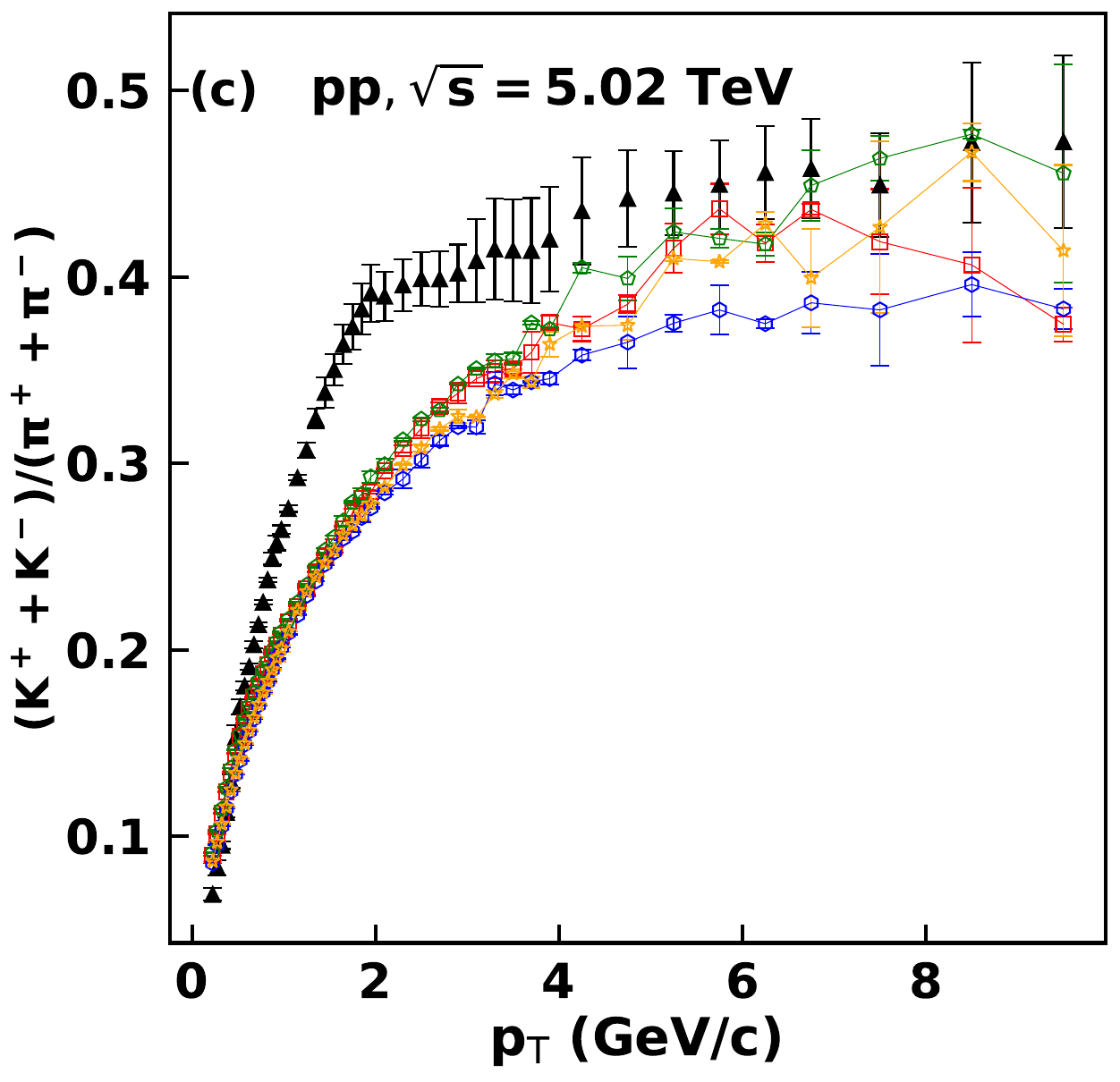}
    \end{overpic}
	}
    \hfill
    \subfloat{
	\begin{overpic}[width=0.32\linewidth]{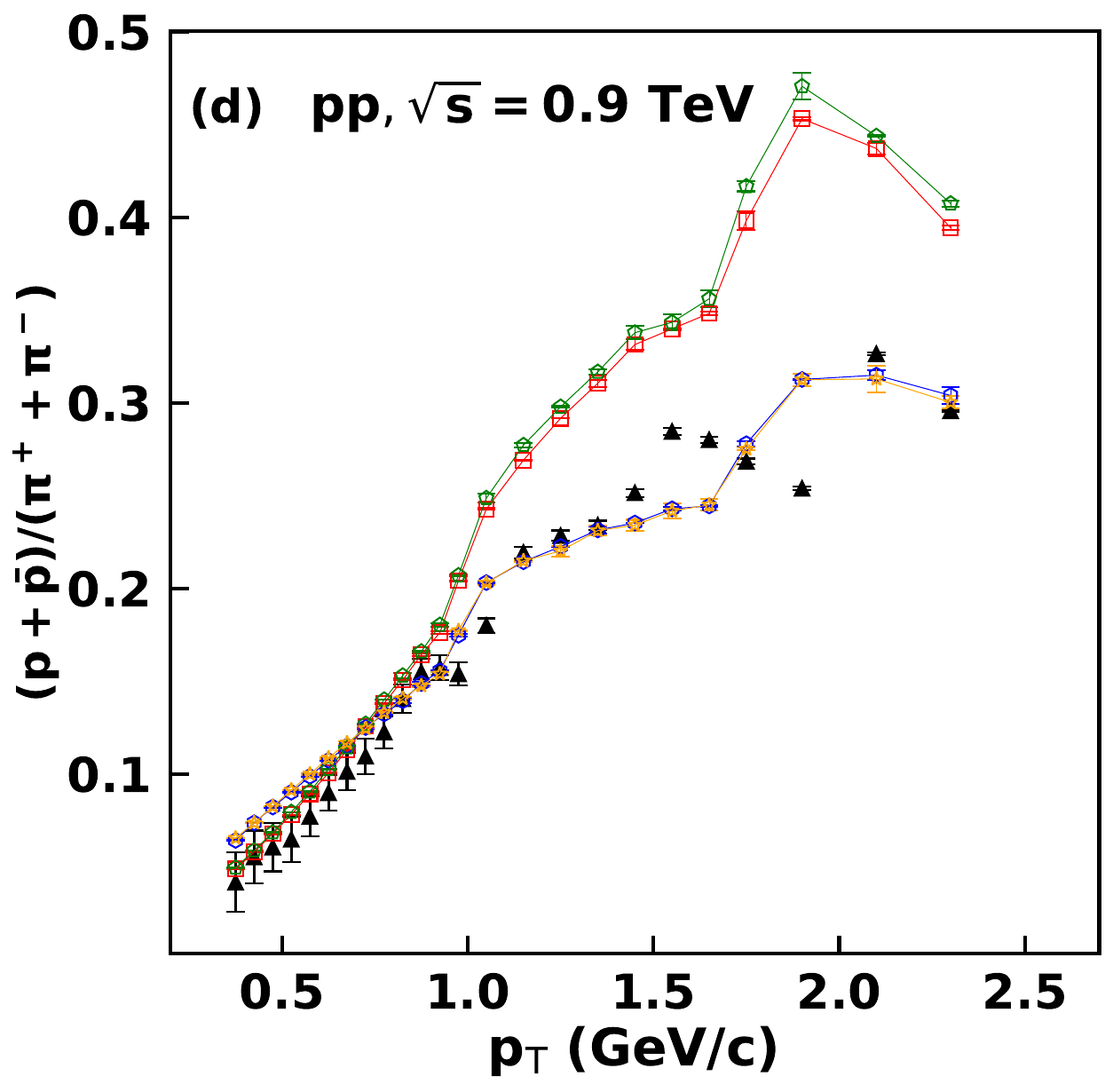}
    \end{overpic}
	}
    \hfill
   \subfloat{
	\begin{overpic}[width=0.32\linewidth]{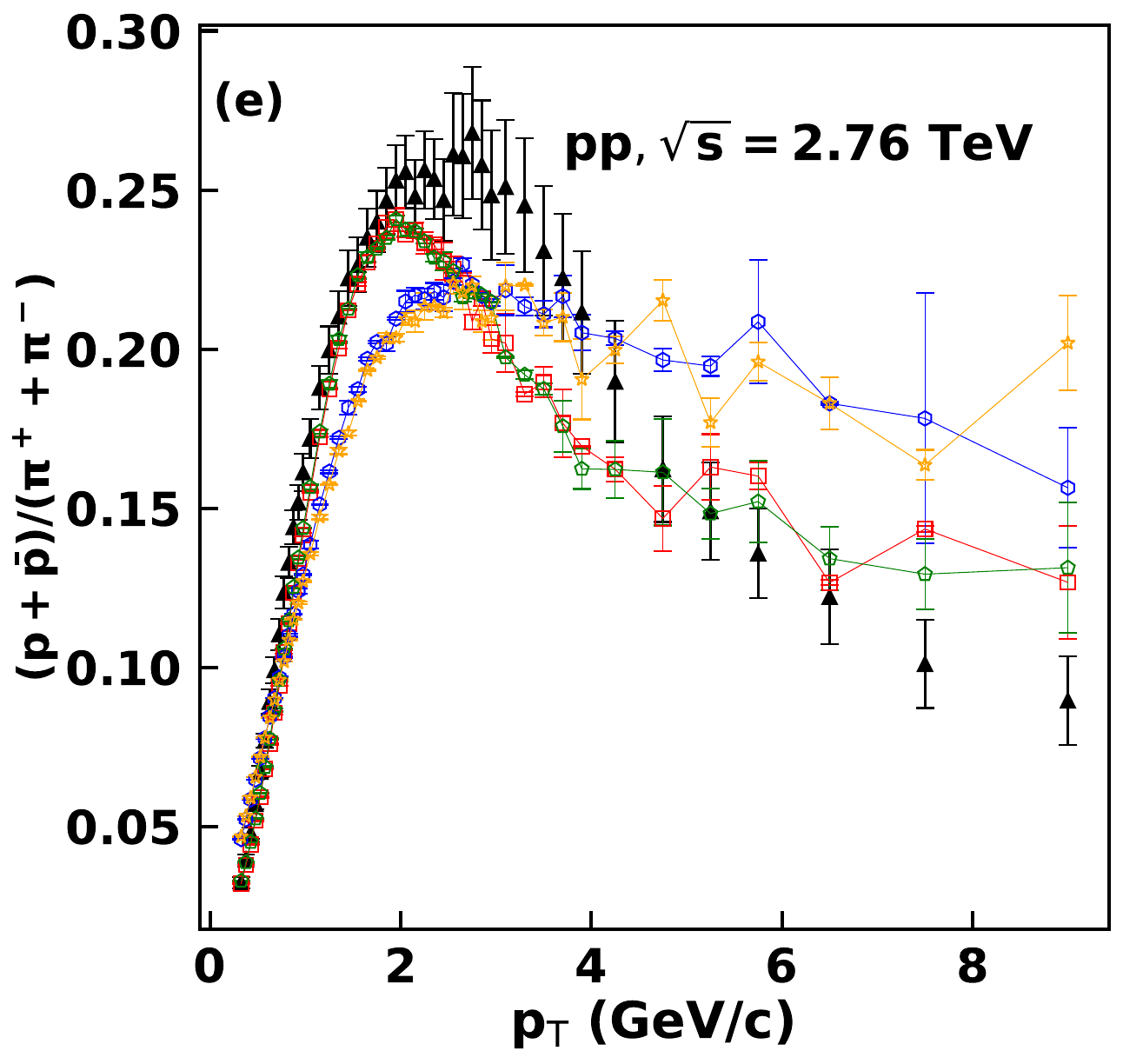}
    \end{overpic}
	}
   \hfill
   \subfloat{
	\begin{overpic}[width=0.32\linewidth]{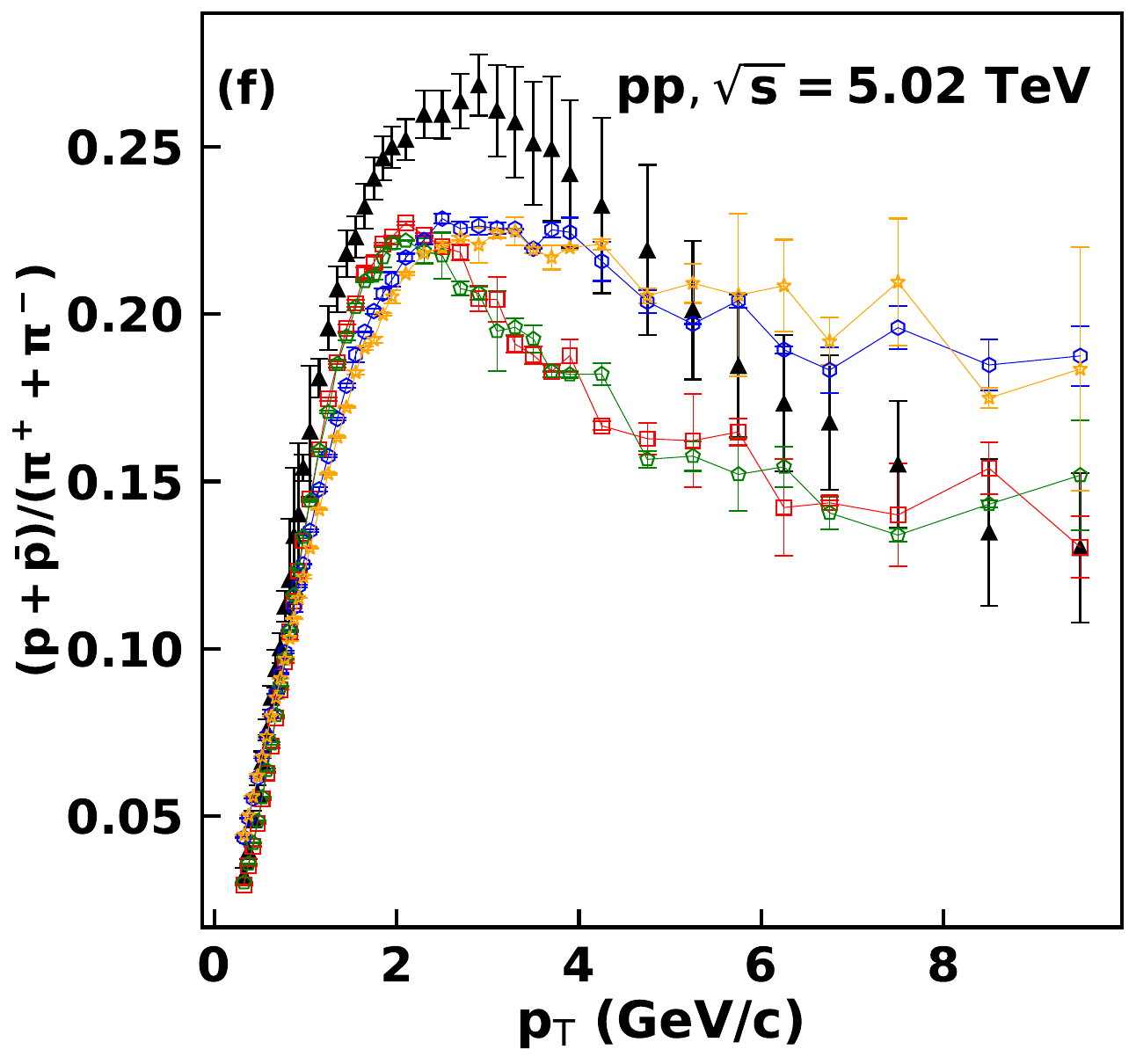}
    \end{overpic}
	}
   \caption{The $K/\pi$ (top) and $p/\pi$ (bottom) ratios as a function of transverse momentum at midrapidity in pp collisions at $\sqrt{s}= 0.9, 2.76$ and $5.02$ TeV from PACIAE 4.0 model simulations compared with ALICE data\cite{ALICE:2011gmo,ALICE:2014juv,ALICE:2019hno}. The error bars for the simulated results are the statistical uncertainties.}
  \label{fig8}
\end{figure*}

To further validate the performance of PACIAE 4.0 model in pp collisions, Fig. \ref{fig8} shows the simulated results of $K/\pi$ and $p/\pi$ particle ratios at collision energies $\sqrt{s}=0.9, 2.76$ and 5.02 TeV. The corresponding experimental results are also shown. Overall, the general trends of particle ratios are well reproduced. Some sensitivities of the particle ratios at high $p_{\rm T}$ to the partonic rescattering and/or hadronic rescattering are observed. 

\section{CONCLUSIONS}\label{con}

With the latest released version of the parton and hadron cascade model PACIAE 4.0, pp collisions at center-of-mass energies ranging from $\sqrt{s}=200$ GeV to 13 TeV are simulated, where the Lund string fragmentation mechanism for the particle production is applied. The effects of the partonic rescattering and hadronic rescattering are also investigated. The results show that the rescattering effects are important to make the simulated identified particle spectra reproducing the experimental data, especially at relative low collision energies. At higher collision energies at LHC, the partonic rescattering is pivotal which increases the total multiplicity of charged particles. The pseudorapidity density distributions of charged particles and the transverse momentum spectra of identified particles have been investigated and compared with the available experimental data. Almost perfect agreement has been achieved which enables us to confidently study the physics in pp collisions with the Monte Carlo approach systematically and provide the baseline for the relativistic heavy-ion collisions. We also can provide the reliable simulation data for the pp collisions at the regions where the experimental data are not available currently. Our results can be a valuable source for the community. More elementary collisions, such as proton-antiproton (p$\bar{\textrm{p}}$) collisions, will be studied in the same framework in a following paper.

\section*{Acknowledgement}
 This work is supported by the National Natural Science Foundation of China under grant Nos. 11447024, 11505108 and 12375135, and by the 111 project of the foreign expert bureau of China. Y.L.Y. acknowledges the financial support from Key Laboratory of Quark and Lepton Physics in Central China Normal University under grant No. QLPL201805 and the Continuous Basic Scientific Research Project (No, WDJC-2019-13). W.C.Z. is supported by the Natural Science Basic Research Plan in Shaanxi Province of China (No. 2023-JCYB-012). H.Z. acknowledges the financial support from Key Laboratory of Quark and Lepton Physics in Central China Normal University under grant No. QLPL2024P01.

\bibliography{reference.bib}

\end{document}